\documentstyle[twoside,fleqn,espcrc2]{article}
\newcommand{\AmS}{{\protect\the\textfont2
  A\kern-.1667em\lower.5ex\hbox{M}\kern-.125emS}}
\def\Tr{{\rm Tr}}

\title{Lattice Effective Actions and Light-Quark Confinement}

\author{K.~Cahill\address{New Mexico Center for Particle Physics,
University of New Mexico, Albuquerque, NM 87131-1156,USA}
        and 
        G.~Herling\address{Center for Advanced Studies, 
	University of New Mexico, Albuquerque, NM 87131-1156,USA}}
       
\begin{document}

\begin{abstract}
The positive-plaquette Manton action at weak coupling
is a reasonable action for short-distance phenomena.
We propose an iterative scheme for evolving 
this action into an effective action
for longer distance scales.
We report on the first step of this scheme
in which we have measured ``blocked'' Creutz ratios
with lattice spacing $ 2a $ at $ \beta = 16 $
on a $ 32^4 $ lattice and have searched
for an effective action that 
yields the same ratios on a $ 16^4 $ lattice.
\par
We also suggest a mechanism for quark confinement
that relies upon the lightness of the $ u $ and $ d $ quarks
and formulate a way of testing it in lattice simulations
of $QCD$.
\end{abstract}

\maketitle

\section{INTRODUCTION}

Most lattice simulations are guided by
Wilson's action~\cite{wilson}
in which the matrices of a compact group
replace the fields of the continuum theory.
In these simulations charged particles
are confined at strong coupling
whether the gauge group is abelian~\cite{creutzab}
or non-abelian~\cite{creutznab,Creu80}.
There have been a few lattice simulations
in which the basic variables are fields.
Some of these non-compact simulations have
no exact gauge symmetry and have shown no sign
of confinement for either abelian
or non-abelian theories~\cite{old,new}.
In others gauge invariance was partially restored
by the imposition of random gauge transformations,
but it is unclear whether the resulting
weak confinement signal~\cite{random}
was due to the decorrelations produced
by the noise of the random gauge transformations
or to the attractive forces of the gauge bosons.
Some very interesting simulations~\cite{palum,cagh}
possess an exact lattice gauge symmetry
and display confinement for $SU(2)$ and $SU(3)$
but not for $U(1)$ above $\beta=0.5$\@.
The gauge fields of these simulations,
however, are not hermitian.
In all simulations, whether compact or non-compact,
confinement has appeared only
when accompanied by significant lattice artifacts.
\par
At small coupling the Wilson action and the Manton action~\cite{manton}
are plausible fundamental actions
suitable for the description of gauge fields
at short distances.
But due to their artifacts, these actions cannot
be fundamental actions at moderate or strong coupling.  
At such couplings these actions might be effective actions 
suitable for describing gauge fields at 
longer distances of the order of a fraction of a fermi. 
Yet it is also possible, as Gribov has suggested~\cite{gribov},
that the mechanism of confinement depends in a crucial way 
upon the lightness of the $ u $  and $ d $  quarks.
In this paper we shall discuss these two possibilities
and shall suggest ways of testing them.

\section{EFFECTIVE ACTIONS}

At very weak coupling,
the best single-plaquette compact action
is probably the one proposed by Manton,
which for an SU(2) plaquette $P$ is 
\begin{equation}
S = \frac{\beta}{2} \arccos^2 \left( \frac{1}{2} \Tr ( P ) \right).
\end{equation}
When the plaquette is close to the identity,
this action is proportional
to the square of the compact field strength
and so resembles the classical action.
At $ \beta < 16 $ it is useful to reject
plaquettes that have negative trace.
\par
To determine what effective action
this fundamental, short-distance action
evolves into at longer distances,
we have begun a simulation of $ SU(2) $
gauge theory at $ \beta= 16 $ on a $ 32^4 $ lattice.
On this lattice our action is Manton's
but with plaquettes of negative trace assigned
infinite action.
We measure Wilson loops $ W(i,j) $ up to $ 12 \times 12 $
and compute both ordinary Creutz ratios and blocked Creutz ratios
\begin{equation}
\chi(i,j,2) = - \log \left( \frac{ W(i,j) W(i-2,j-2) }
{ W(i-2,j) W(i,j-2) } \right)
\end{equation}
in which the lattice spacing is $ 2a $ instead of $ a $
and both $ i $ and $ j $ are even. 
Our strategy is to experiment with arbitrary actions
on $ 16^4 $ lattices so as to find one whose
ordinary Creutz ratios are equal to the blocked Creutz ratios
on the $ 32^4 $ lattice.
If we find such an effective action,
then we plan to use it on a new $ 32^4 $ lattice
and to measure both ordinary and blocked Creutz ratios
on that lattice.
The next step would be to search for a second effective
action that on a second $ 16^4 $ lattice gives
ordinary Creutz ratios that are equal to the blocked
ones of the second $ 32^4 $ lattice.
\par
The principal problem with this scheme is that 
large Wilson loops converge very slowly at weak coupling.
In the first step of our implementation of this procedure
at $ \beta = 16 $, we performed 392,000 thermalization sweeps
on a $ 16^4 $ lattice and then used 16 clones of this lattice as
our initial $ 32^4 $ lattice.
After 20,000 thermalization sweeps
on the $ 32^4 $ lattice,
the blocked Creutz ratios
$ \chi(4,4,2) = 0.01531(6) $,
$ \chi(4,6,2) = 0.01035(6) $, and
$ \chi(4,8,2) = 0.00950(5) $ 
of the smaller loops may be close to converging.
But the blocked ratios 
$ \chi(6,6,2) = 0.00465(4) $,
$ \chi(6,8,2) = 0.00354(4) $, and
$ \chi(8,8,2) = 0.00227(5) $ 
of the larger loops are still trending upward.
The errors quoted for these large loops are purely
statistical and do not contain the systematic error
of the secular drift.  We have searched for an effective action
that would yield these ratios on a $ 16^4 $ lattice;
the closest so far is one with $ \beta = 15 $.
\par
One may iterate this scheme provided one can find
a new effective action at each step.
In this case the successive effective actions
eventually should evolve into a suitable
long-distance effective action.
This action might indeed be turn out to be
Wilson's action at moderate coupling.
After a few iterations, however,
there may be no available single-plaquette effective action
that yields Creutz ratios on the $ 16^4 $ lattice
that are approximately equal to the blocked ratios 
of the preceding $ 32^4 $ lattice.
If the scheme hits a wall in this way,
then present lattice methods may be meaningless,
and the reason for confinement may be the 
lightness of the lighter quarks. 

\section{LIGHT-QUARK CONFINEMENT}

We now wish to propose a simple dynamical
mechanism for quark confinement 
which implements Gribov's idea~\cite{gribov}
about the possible importance of light quarks.
We also shall suggest a way of testing this
mechanism.  
\par
The hamiltonian of $ QCD $
describing the interaction of the light
$ u $ and $ d $ quarks with the $ SU(3) $
gauge fields $ A_\mu^b $
contains the pieces
\begin{equation}
V = - i g \, \bar u \not \!\! A^b t_b \, u
- i g \, \bar d \not \!\! A^b t_b \, d 
\end{equation}
and
\begin{equation}
M = m_u \, \bar u \, u + m_d \, \bar d \, d .
\end{equation} 
Because $ V $ does not have a definite sign,
it seems possible that in the physical vacuum of $ QCD $
the light quarks $ u $ and $ d $ might condense in ways that
are correlated with the fluctuating gauge field $ A^b_\mu $.
In this picture
the physical vacuum is represented by a functional integral
\begin{equation}
| \Omega \rangle = \int D A^b_\mu \Psi( A^b_\mu, u, d ) 
\, | A^b_\mu, u, d \rangle
\end{equation}
over a state $ | A^b_\mu, u, d \rangle $ 
in which the gluon variables form
something like a coherent state 
with mean value $ A^b_\mu(x) $ and 
in which the quark variables $ u $ and $ d $ 
are correlated with the field $ A^b_\mu(x) $
in such a way that
the mean value of the interaction $ V $ 
is large and negative
\begin{equation} 
\langle \Omega | V | \Omega \rangle < 0 .
\end{equation}
Because the $ u $ and $ d $ quarks are light,
the effect $ \langle \Omega | M | \Omega \rangle $
of their masses is small,
and the mean value of $ V + M $ is large and negative 
\begin{equation}
\langle \Omega | V + M | \Omega \rangle < 0 .
\end{equation}
\par
In such a physical vacuum,
pairs of up and down quarks from the condensate 
can convert pairs of quarks created in the debris
of hadronic collisions into mesons. 
\par
Inasmuch as the proposed confinement mechanism
is intrinsically non-perturbative,
it is probably necessary to test it in
a lattice simulation.
The signal would be a drop in the euclidean action
$ S_q[U,u,d, \bar u, \bar d] $ of the quarks
with the onset of confinement, 
as indicated by the Creutz ratios of the Wilson loops.
We intend to use the QCDF90 codes~\cite{rebbi}
to perform this test.

\section{ACKNOWLEDGEMENTS} 

We are grateful Michael Creutz for helpful 
advice, to Claudio Rebbi and Andrea Ruben Levi
for sending us QCDF90, and to Dmitri Sergatskov
for help in setting up Linux on a multi-pentium system.
Research sponsored in part by the Phillips Laboratory, 
Air Force Materiel Command, USAF, under cooperative
agreement number F29601-93-2-0001. 
The views and conclusions contained in this document are those of the
authors and should not be interpreted as necessarily representing 
the official policies or endorsements, 
either expressed or implied, of Phillips Laboratory or the U.S. Government.
%appear at the top of a page).
%
%\subsection{Tables}
%
%Tables should be presented in the form shown in

%References should be collected at the end of your paper. Do not begin
%them on a new page unless this is absolutely necessary. They should be
%prepared according to the sequential numeric system making sure that
%all material mentioned is generally available to the reader. Use
%\verb+\cite+ to refer to the entries in the bibliography so that your
%accumulated list corresponds to the citations made in the text body. 

\end{document}